# Dynamic heart rate variability for classification of arrhythmia burden

Brian Kang[1], Michael Wacker[1], Steven S. Kang[1]*

[1]University of Missouri-Kansas City, Missouri, United States

* kangseung@umkc.edu

## ABSTRACT

**Purpose**: Heart rate variability (HRV) may provide information beyond heart rate (HR) for identifying arrhythmia burden, but the value of temporal HRV variation remains unclear. This study developed a dynamic HRV measure quantifying the interquartile range of window-level root mean square of successive RR-interval differences ($IQR_{RMSSD}$) and evaluated its utility in classifying total abnormal-beat and premature ventricular contraction (PVC) burden compared with the conventional RR-based HRV measure, RMSSD. **Methods**: Manually annotated 30-minute ECG recordings from 43 individuals were analyzed using overlapping windows. A linear support vector classifier was used to classify individuals as having high or low arrhythmia burden, defined using 1% and 10% burden thresholds. Four feature sets were evaluated: HR alone, HR with RMSSD, HR with $IQR_{RMSSD}$, and all three measures. Performance was assessed using repeated stratified five-fold cross-validation. **Results**: HRV-containing feature sets generally outperformed HR alone, depending on outcome and threshold. At the 1% threshold, $IQR_{RMSSD}$ showed particularly high sensitivity for total abnormal-beat and PVC burden (0.970 ± 0.064 and 0.882 ± 0.193, respectively). At the 10% threshold, conventional RMSSD showed the strongest overall discrimination, achieving a ROC-AUC of 0.947 for PVC burden. Combined features sometimes improved the sensitivity-specificity balance but not overall discrimination. **Conclusion**: These findings suggest that a conventional RR-based HRV measure is informative for elevated ectopic burden, while a dynamic HRV measure may provide complementary information for lower-burden abnormalities.

## INTRODUCTION

Heart rate variability (HRV) is a widely used noninvasive measure of cardiac autonomic regulation, reflecting beat-to-beat variation in the intervals between normal heartbeats [1]. Among commonly used HRV indices, the root mean square of successive differences between normal-to-normal intervals (RMSSD) is frequently used to quantify short-term variability and is considered particularly sensitive to parasympathetic cardiac modulation [2]. Thus, HRV offers clinically meaningful information beyond mean heart rate (HR) alone and is useful for evaluating cardiac regulatory function and disease-related alterations.

A growing body of clinical literature suggests that HRV-derived measures may provide useful information for predicting new-onset atrial fibrillation (AF) and recurrence after AF treatment, although their clinical utility remains under investigation [3,4]. In the population-based Oulu Project Elucidating Risk of Atherosclerosis (OPERA) study, reduced low-frequency HRV, reflecting combined sympathetic and parasympathetic modulation, was independently associated with increased risk of symptomatic AF [5]. In perioperative settings, nonlinear HRV indices such as detrended fluctuation analysis alpha 1, which reflects short-term fractal organization of beat-to-beat intervals, have also shown moderate value for predicting postoperative AF after coronary artery bypass grafting [6].

Distinguishing low- from high-burden arrhythmia groups is clinically important because arrhythmia burden is increasingly recognized as a quantitative marker of disease severity, prognosis, and treatment response rather than a simple indicator of arrhythmia presence. Greater AF burden has been associated with adverse cardiovascular outcomes [7,8], while high premature ventricular contraction (PVC) burden is associated with PVC-induced cardiomyopathy and can inform decisions about additional monitoring or rhythm-directed intervention [9].

Conventional cardiac measures, including mean HR and standard HRV indices such as RMSSD, summarize overall cardiac activity and beat-to-beat variability over a fixed recording period. This static approach can reduce complex temporal patterns to a single global value and may be insufficient in assessing arrhythmia, where intermittent rhythm disturbances, frequent ectopic beats, or clustered events produce substantial temporal fluctuations. Two individuals may therefore show similar average HR or overall RMSSD while differing markedly in the temporal instability of their RR-interval patterns. Consequently, static HRV measures may miss clinically relevant fluctuations associated with arrhythmia burden [10–12].

Dynamic HRV analysis may provide a solution by calculating HRV within moving time windows across an ECG recording. Zhang and colleagues previously investigated "momentary," aspects of HRV using overlapping windows to track changes in conventional measures [13]. Using a related approach, we developed a novel dynamic HRV metric that quantifies temporal variability in short-term RMSSD. To our knowledge, this is the first dynamic HRV measure based on the dispersion of short-window RMSSD values across a recording. The resulting measure, $IQR_{RMSSD}$, is a second-order measure of first-order RR-based HRV, defined as the interquartile range (IQR) of windowed RMSSD values. Conceptually, it captures temporal instability rather than the average magnitude of HRV alone. Although prior studies have applied machine-learning methods to HRV-derived features for segment-level arrhythmia detection, particularly AF [14,15], relatively few have examined whether HRV features can classify individuals according to overall arrhythmia burden. We therefore tested whether $IQR_{RMSSD}$ provides predictive information beyond conventional RMSSD for distinguishing low versus high arrhythmia burden using parsimonious, interpretable support vector machine (SVM) models.

## METHODS

### Dataset and ECG Recordings

We used ECG data from the MIT-BIH Arrhythmia Database [16]. The database comprises 48 approximately 30-min recordings from 47 individuals, including 25 men aged 32–89 years and 22 women aged 23–89 years. Because two recordings originated from the same individual, one was removed to ensure independence among subject-level observations. Four recordings with most beats generated by pacemakers were also excluded, yielding 43 recordings. ECG signals were obtained from two-channel ambulatory recordings and digitized at 360 Hz with 11-bit resolution. Modified limb lead II (MLII) was available for most recordings and was used in the present analysis. Beat locations and types were independently annotated by expert cardiologists, with disagreements resolved to establish the reference annotations. The dataset and documentation are publicly available through PhysioNet [16], with a details provided by Moody and Mark [17].

### ECG preprocessing and burden calculation

The first two beat annotations of each recording were removed because they lacked complete RR-interval information. Analyses were restricted to predefined beat classes: atrial premature (A), aberrated atrial premature (a), premature ventricular contraction (V), ventricular escape (E), nodal/junctional premature (J), supraventricular premature/ectopic (S), fusion (F), left bundle branch block (L), right bundle branch block (R), normal (N), and unclassifiable (Q). RR intervals associated with excluded or invalid labels were removed, and beat occurrence times were converted to milliseconds to generate a continuous beat-time series.

Beat annotations were used to derive two arrhythmia-burden measures across each approximately 30-min recording. Total abnormal-beat burden was calculated as the proportion of analyzed beats labeled A, a, V, E, J, S, or F. PVC burden included only V beats. These outcomes therefore represented overall included abnormal-beat burden and PVC-specific burden, respectively.

$$\text{Burden (\%)} = \frac{\text{Number of Target Beats}}{\text{Number of Valid Annotated Beats}} \times 100$$

For each outcome, subjects were classified into low- and high-burden groups using prespecified 1% and 10% thresholds of total analyzed beats. High burden was defined as a proportion greater than the threshold and low burden as a proportion at or below it. Prior studies identify PVC burden of approximately 10% or greater as clinically meaningful because of associations with left ventricular dysfunction and reversible PVC-induced cardiomyopathy [18]. The 1% threshold was selected as an exploratory cutoff representing relatively low ectopic burden [19].

### HRV Measures

Mean HR was calculated from each subject's mean RR interval (HR = 60/mean RR, with RR in seconds). Although conventional HRV analysis typically uses normal-to-normal (NN) intervals, we retained ectopic-associated RR intervals because our goal was to characterize arrhythmia-related variability rather than isolated autonomic modulation. Prior arrhythmia-focused studies have similarly used RR or heart-period information to detect cardiac abnormalities [20]. Excluding ectopic intervals could therefore remove variability directly related to the arrhythmic events of interest. RMSSD was calculated from successive RR-interval differences as follows:

$$\mathrm{RMSSD}=\sqrt{\frac{1}{\mathrm{N}-1}\sum_{\mathrm{i}=1}^{\mathrm{N}-1}\left(\mathrm{RR}_{\mathrm{i}+1}-\mathrm{RR}_{\mathrm{i}}\right)^{2}}$$

RMSSD was selected because it can be reliably estimated from short recordings and is therefore well suited for moving-window analysis; many frequency-domain HRV measures require longer intervals for stable estimation. The short window was used to obtain repeated HRV estimates within each approximately 30-min recording while retaining sufficient beats for RMSSD calculation. RMSSD was first calculated across the complete RR time series. To characterize temporal variation, each recording was then divided into 60-s windows advanced in 30-s steps (50% overlap; **Figure 1**). A 60-s window was selected because RMSSD estimated from 1-min ECG segments shows good agreement with conventional 5-min RMSSD [21]. RMSSD was calculated for each window across the recording.

$$\mathrm{IQR}_{\mathrm{RMSSD}} = \mathrm{Q3(RMSSD)} - \mathrm{Q1(RMSSD)}.$$

$\mathrm{IQR}_{\mathrm{RMSSD}}$ was calculated for each subject from the resulting window-level RMSSD distribution. The IQR was selected as a robust, nonparametric measure of dispersion because it summarizes the central 50% of observations and is less sensitive than standard deviation to isolated extreme values and skewed distributions. Thus, $\mathrm{IQR}_{\mathrm{RMSSD}}$ represents the typical range of short-term HRV fluctuation across a recording. **Figure 1** summarizes the ECG-to-feature workflow and illustrates how conventional RMSSD and $\mathrm{IQR}_{\mathrm{RMSSD}}$ represent different aspects of RR variability.

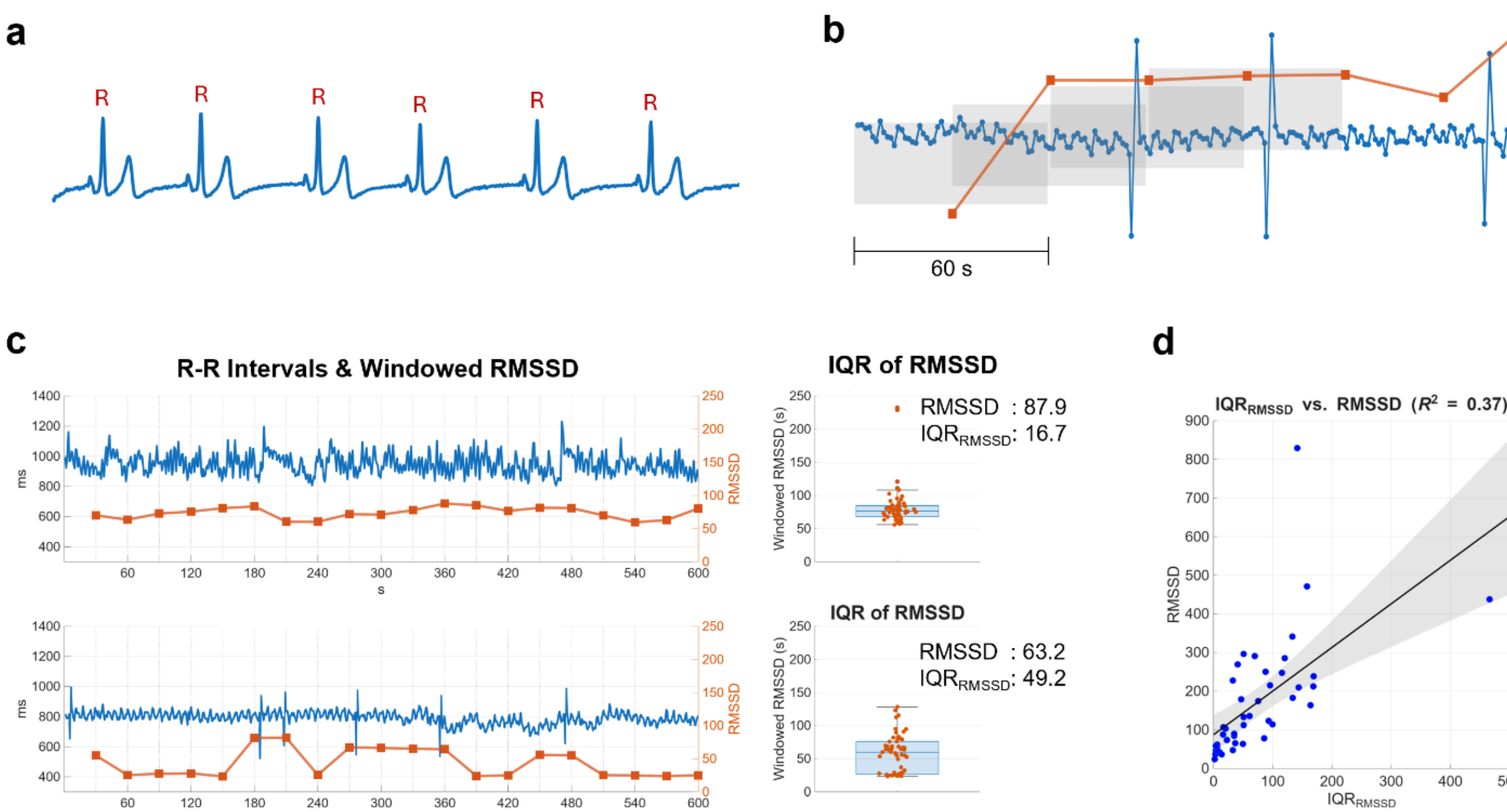

**Figure 1.** Overview of dynamic HRV derivation and arrhythmia-burden classification. (a) RR intervals from approximately 30-min ECG recordings. (b) RMSSD calculated in 60-s windows advanced by 30 s. (c) Windowed RMSSD and $\mathrm{IQR}_{\mathrm{RMSSD}}$ in two individuals illustrate distinct properties. The bottom individual had slightly lower RMSSD but markedly higher $\mathrm{IQR}_{\mathrm{RMSSD}}$, reflecting greater short-term HRV fluctuations. (d) RMSSD and $\mathrm{IQR}_{\mathrm{RMSSD}}$ were significantly but moderately correlated ($r = 0.61$; $R^2 = 0.37$), indicating substantial nonshared variance.

### Machine Learning Modeling

We developed four classification models using distinct feature sets: (1) HR-only reference model; (2) first-order model (HR + RMSSD); (3) second-order model (HR + $IQR_{RMSSD}$); and (4) combined model (HR + RMSSD + $IQR_{RMSSD}$). This structure assessed the incremental information provided by static and dynamic HRV beyond average HR. RMSSD and $IQR_{RMSSD}$ were not corrected for prevailing HR; instead, HR was included in each learned feature set to account for between-subject differences in mean HR, although residual confounding remains possible. Analyses were conducted in Python using WFDB for ECG/annotation processing, NumPy and Pandas for data manipulation, SciPy for HRV calculations, and scikit-learn for modeling. A fixed random seed was used for stochastic procedures to support reproducibility.

Classification was performed with a Support Vector Classifier (SVC) using a linear kernel, selected to limit model complexity given the small sample and limited predictors. A stratified dummy classifier served as a chance-level benchmark by randomly predicting classes according to training-set class proportions. To address class imbalance, the SVC used inverse-frequency class weights (class_weight='balanced'), assigning greater weight to minority-class errors and lower weight to majority-class errors during training.

Predictors were standardized using z-scores within the modeling pipeline so that scaling parameters were estimated only from training data in each fold, preventing information leakage. Repeated stratified five-fold cross-validation with 10 repetitions (n_splits=5, n_repeats=10) was used to approximately preserve class proportions within folds and reduce variability in performance estimates given the limited sample size.

### Machine Learning Model Performance Evaluation

Model performance was assessed using sensitivity, specificity, balanced accuracy, Matthews correlation coefficient (MCC), and receiver operating characteristic area under the curve (ROC-AUC). Sensitivity was considered particularly important because it quantifies the proportion of high-burden cases correctly identified, whereas specificity reflects correct identification of low-burden cases. Balanced accuracy averages sensitivity and specificity, MCC summarizes all four confusion-matrix outcomes and is robust to class imbalance, and ROC-AUC measures discrimination across decision thresholds. Metrics were reported as mean ± standard deviation across cross-validation folds; 95% empirical intervals for ROC-AUC were estimated from the 2.5th and 97.5th percentiles.

Comparisons among feature sets were primarily descriptive because of the limited sample and dependence among repeated cross-validation folds; differences in ROC-AUC and other performance measures were emphasized rather than formal hypothesis testing. Models were evaluated at 1% and 10% thresholds for total abnormal-beat and PVC burden. Robustness analyses examined alternative window overlaps (0% and 25% versus the primary 50%), logistic regression and random forest classifiers, and HRV features recalculated from consecutive non-ectopic intervals. Alternative classifiers were treated as secondary robustness analyses rather than a basis for selecting the best-performing algorithm.

## RESULTS

The number of arrhythmic beats varied substantially among subjects (**Table 1**). At the 1% threshold, PVC classification was nearly balanced (21 high versus 22 low burden), whereas the 10% threshold was less balanced (10 versus 33). Total abnormal-beat classification included 30 high versus 13 low subjects at 1% and 14 versus 29 at 10%, providing different degrees and directions of class imbalance across outcomes.

**Table 1.** Types of arrhythmia and their descriptive statistics

| Type of Arrhythmia | Mean | Standard Deviation | Range |
|---|---|---|---|
| Total abnormal beats | 240.7 | 335.0 | 0–1382 |
| Premature Ventricular Contractions | 155.9 | 250.7 | 0–992 |
| Ventricular Arrhythmic beats | 159.3 | 248.0 | 0–992 |
| Supraventricular Arrhythmic beats | 63.2 | 212.4 | 0–1382 |

As summarized in **Tables 2** and **3**, HRV-based models substantially outperformed HR alone. At the 1% total abnormal-beat threshold, the second-order model achieved the highest sensitivity (0.970 ± 0.064), balanced accuracy (0.910 ± 0.109), and MCC (0.846 ± 0.190), identifying nearly all high-burden subjects while maintaining strong specificity (0.850 ± 0.198). Its ROC-AUC was slightly lower than that of the first-order model (0.919 ± 0.159 vs. 0.926 ± 0.099). The combined model improved specificity to 0.900 ± 0.183 but reduced sensitivity and MCC relative to the second-order model, indicating no clear advantage from combining RMSSD and $IQR_{RMSSD}$ at this threshold.

At the 10% total abnormal-beat threshold, all HRV-based models again outperformed HR alone. In contrast to the 1% analysis, the first-order model showed the strongest overall performance, with sensitivity 0.813 ± 0.218, specificity 0.817 ± 0.187, MCC 0.644 ± 0.248, and ROC-AUC 0.933 ± 0.092. The second-order and combined models were less informative than RMSSD at this higher threshold, supporting a threshold-dependent difference between static and dynamic HRV.

**Table 2.** Classification performance for total abnormal-beat burden

| Feature Set | Sensitivity | Specificity | Balanced Accuracy | MCC | ROC-AUC |
|---|---|---|---|---|---|
| 1% Threshold | | | | | |
| Baseline (dummy) | 0.547 | 0.547 | 0.547 | 0.000 | 0.547 |
| HR only | 0.600 (0.203) | 0.683 (0.295) | 0.642 (0.156) | 0.278 (0.294) | 0.731 (0.205) |
| HRV (HR + RMSSD) | 0.730 (0.170) | 0.887 (0.187) | 0.808 (0.127) | 0.589 (0.241) | 0.926 (0.099) |
| Dynamic HRV (HR + $IQR_{RMSSD}$) | 0.970 (0.064) | 0.850 (0.198) | 0.910 (0.109) | 0.846 (0.190) | 0.919 (0.159) |
| Combined (HR + RMSSD + $IQR_{RMSSD}$) | 0.877 (0.156) | 0.900 (0.183) | 0.888 (0.120) | 0.769 (0.230) | 0.925 (0.144) |
| 10% Threshold | | | | | |
| Baseline (dummy) | 0.460 | 0.460 | 0.460 | 0.000 | 0.460 |
| HR only | 0.520 (0.223) | 0.643 (0.168) | 0.581 (0.147) | 0.161 (0.287) | 0.674 (0.180) |
| HRV (HR + RMSSD) | 0.813 (0.218) | 0.817 (0.187) | 0.815 (0.128) | 0.644 (0.248) | 0.933 (0.092) |
| Dynamic HRV (HR + $IQR_{RMSSD}$) | 0.743 (0.263) | 0.689 (0.198) | 0.716 (0.122) | 0.436 (0.242) | 0.815 (0.120) |
| Combined (HR + RMSSD + $IQR_{RMSSD}$) | 0.763 (0.269) | 0.813 (0.191) | 0.788 (0.148) | 0.589 (0.291) | 0.919 (0.108) |

For PVC burden at 1%, HR alone provided moderate discrimination (ROC-AUC 0.707 ± 0.170) but weak sensitivity (0.583 ± 0.211). The first-order model achieved the highest ROC-AUC (0.821 ± 0.139) and specificity (0.749 ± 0.185). The second-order model showed slightly lower ROC-AUC (0.800 ± 0.131) but greater sensitivity (0.882 ± 0.193) and MCC (0.560 ± 0.249), at the cost of lower specificity (0.645 ± 0.186). The combined model produced the highest balanced accuracy (0.788 ± 0.125) and MCC (0.595 ± 0.250) while retaining high sensitivity, but its ROC-AUC did not exceed either individual HRV model.

At the 10% PVC threshold, the first-order model clearly demonstrated the strongest performance, with ROC-AUC 0.947 ± 0.083, MCC 0.714 ± 0.212, sensitivity 0.950 ± 0.150, and specificity 0.823 ± 0.148. The combined model also performed strongly, but adding $IQR_{RMSSD}$ only slightly increased specificity (0.847) while reducing ROC-AUC, MCC, and sensitivity. The second-order model showed substantially lower discrimination and sensitivity. Thus, $IQR_{RMSSD}$ provided no clear incremental value for PVC burden above 10%.

Sensitivity analyses for the 1% total abnormal-beat outcome indicated generally stable performance across window-overlap settings. For the second-order model, ROC-AUC ranged from 0.914 to 0.919, MCC from 0.757 to 0.846, and sensitivity from 0.907 to 0.970 across 0%, 25%, and 50% overlap.

**Table 3.** Classification performance for PVC burden

| Feature Set | Sensitivity | Specificity | Balanced Accuracy | MCC | ROC-AUC |
|---|---|---|---|---|---|
| 1% Threshold | | | | | |
| Baseline (dummy) | 0.452 | 0.452 | 0.452 | 0.000 | 0.452 |
| HR only | 0.583 (0.211) | 0.679 (0.256) | 0.631 (0.145) | 0.280 (0.306) | 0.707 (0.170) |
| HRV (HR + RMSSD) | 0.760 (0.200) | 0.749 (0.185) | 0.754 (0.120) | 0.528 (0.241) | 0.821 (0.139) |
| Dynamic HRV (HR + $IQR_{RMSSD}$) | 0.882 (0.193) | 0.645 (0.186) | 0.763 (0.125) | 0.560 (0.249) | 0.800 (0.131) |
| Combined (HR + RMSSD + $IQR_{RMSSD}$) | 0.842 (0.179) | 0.735 (0.187) | 0.788 (0.125) | 0.595 (0.250) | 0.793 (0.136) |
| 10% Threshold | | | | | |
| Baseline (dummy) | 0.517 | 0.517 | 0.517 | 0.000 | 0.517 |
| HR only | 0.550 (0.320) | 0.630 (0.195) | 0.590 (0.180) | 0.159 (0.326) | 0.667 (0.220) |
| HRV (HR + RMSSD) | 0.950 (0.150) | 0.823 (0.148) | 0.887 (0.104) | 0.714 (0.212) | 0.947 (0.083) |
| Dynamic HRV (HR + $IQR_{RMSSD}$) | 0.530 (0.366) | 0.690 (0.202) | 0.610 (0.185) | 0.199 (0.349) | 0.767 (0.164) |
| Combined (HR + RMSSD + $IQR_{RMSSD}$) | 0.900 (0.200) | 0.847 (0.133) | 0.873 (0.122) | 0.701 (0.237) | 0.930 (0.096) |

MCC and sensitivity were modestly lower without overlap but nearly identical at 25% and 50%. Results were also broadly consistent when the linear SVC was replaced with logistic regression or random forest: across the three classifiers, the second-order feature set achieved ROC-AUC values of 0.895–0.934, MCC values of 0.723–0.846, and sensitivity values of 0.873–0.970. Although the best feature set varied by classifier, RR-variability models consistently outperformed HR alone. Lastly, recalculating features from consecutive non-ectopic intervals (i.e., NN-based HRV measures) reduced performance relative to the primary RR-based models.

## DISCUSSION

This study developed $IQR_{RMSSD}$ as a dynamic HRV measure and investigated its potential utility for classifying arrhythmia burden. As illustrated in **Figure 1**, short-term windowed HRV varied substantially over time, and this variability was captured by $IQR_{RMSSD}$ as a property distinct from conventional HRV. Although developed independently of prior work [13], our approach shares the concept of applying HRV analysis to moving windows. It extends that approach by summarizing the distribution of window-level RMSSD values across an entire recording rather than using the windows only to track moment-to-moment changes. Specifically, $IQR_{RMSSD}$ quantifies the spread of the middle 50% of window-level RMSSD values, providing a measure of the typical range of short-term HRV fluctuations while minimizing the influence of isolated extreme windows.

Conventional RR-based RMSSD summarizes the overall magnitude of successive RR-interval variation, whereas $IQR_{RMSSD}$ describes the extent to which short-term RMSSD varies within a recording. Thus, two individuals with similar conventional RMSSD may differ in the temporal dispersion of their short-term variability. The classification analyses tested whether this dynamic information provides discriminatory value beyond HR and conventional RMSSD.

Across both outcomes and thresholds, models incorporating HRV features consistently outperformed HR alone, but the relative contribution of static and dynamic HRV varied by burden threshold. $IQR_{RMSSD}$ showed its greatest utility when high arrhythmia burden was defined using the 1% threshold, particularly through greater sensitivity in identifying subjects above this cutoff, whereas conventional RMSSD was more informative when the 10% threshold was used and produced the strongest PVC-burden classification. This threshold-dependent pattern suggests that dynamic HRV may be particularly informative when distinguishing high- versus low-burden groups using a relatively low classification threshold, whereas conventional RMSSD may better discriminate groups defined by a higher burden threshold. Greater sensitivity at the 1% threshold indicates that dynamic HRV is better suited to identifying individuals with modest but non-negligible arrhythmic burden who would not meet a higher-burden criterion.

Importantly, low arrhythmia burden should not be interpreted as low overall cardiovascular risk, because reduced conventional HRV, particularly when assessed from NN intervals, has been associated with coronary artery disease, heart failure, and adverse cardiovascular outcomes [22,23]. Conversely, in our ectopy-inclusive analysis, higher RMSSD and $IQR_{RMSSD}$ can reflect greater arrhythmia-related RR irregularity. The models should therefore be interpreted specifically as classifiers of arrhythmia burden rather than general indicators of cardiac health. Combining RMSSD and $IQR_{RMSSD}$ sometimes improved the sensitivity-specificity balance but did not consistently exceed the strongest individual model. Their correlation ($r = 0.61$) suggests overlapping information and potential multicollinearity, which may limit additional predictive gain [24].

Sensitivity analyses indicated that the principal findings were not dependent on a particular overlap percentage or classifier. The similar performance with 25% and 50% overlap suggests that moderate changes in window configuration had little effect on the findings, although these analyses do not establish an optimal overlap. Performance declined when features were recalculated using consecutive non-ectopic intervals, indicating that ectopic-associated RR irregularity contributed substantially to discrimination. Accordingly, the primary RMSSD and $IQR_{RMSSD}$ features should be interpreted as measures of minimally processed RR variability rather than isolated autonomic modulation. This distinction does not invalidate their use for burden classification because arrhythmia-related timing variability was intentionally preserved. The deliberately simple linear SVM was used to evaluate the information contained in the new feature rather than to optimize prediction through more complex algorithms.

Several limitations should be considered. First, the analysis included only 43 recordings from a single public database, limiting precision and generalizability, particularly at the 10% thresholds where high-burden groups were small. Repeated stratified cross-validation and class weighting address some concerns but cannot replace independent validation. Second, the approximately 30-min ECG segments provide short-term burden estimates rather than the 24-h burden commonly measured with Holter monitoring. Because arrhythmias may occur intermittently, burden measured during a short segment may not represent a subject's daily burden or clinical risk. Although these shorter recordings enabled evaluation of dynamic HRV within limited ECG segments, they restrict clinical generalizability. Third, the 1% threshold has less established clinical support than the 10% PVC-burden threshold and should be considered exploratory, particularly for the broader total abnormal-beat outcome. Finally, RMSSD generally declines with age [25], and the wide age range could have confounded associations; the sample was too small for reliable age-adjusted modeling or inclusion of additional demographic covariates. Future studies should evaluate these features in larger, clinically characterized cohorts with longer recordings and independent validation to determine reproducibility, generalizability, and clinical value.

## CONCLUSION

The ability to distinguish individuals with greater arrhythmia burden may have clinical relevance for assessing disease severity and the need for additional monitoring. The proposed $IQR_{RMSSD}$ measure captures within-recording temporal variability in short-term HRV and showed complementary classification value to conventional RMSSD. Its greater sensitivity at the 1% threshold suggests particular utility for identifying individuals above a relatively low arrhythmia-burden threshold, whereas conventional RMSSD performed better at the 10% threshold. These findings support further evaluation of dynamic HRV as a complementary marker rather than a replacement for conventional HRV. Larger independent studies with longer recordings are needed to establish reliability, prognostic value, and clinical utility.

**Funding Statement**

This work was supported by the University of Missouri-Kansas City through the Sarah Morrison Pilot Research Fund awarded to S.K.

**Data Availability**

The data analyzed in this study are publicly available from the MIT-BIH Arrhythmia Database on PhysioNet (version 1.0.0; https://doi.org/10.13026/C2F305).